\documentclass{aastex}
\usepackage{emulateapj5}

\newif\ifAMStwofonts
\AMStwofontstrue

\def\kms{km~s$^{-1}$}

\def\degree{$^{\circ}$}

\def\ga{\mathrel{\hbox{\rlap{\hbox{\lower4pt\hbox{$\sim$}}}\hbox{$>$}}}}
\def\la{\mathrel{\hbox{\rlap{\hbox{\lower4pt\hbox{$\sim$}}}\hbox{$<$}}}}

\def\Nperp {N_{\perp}}

\shorttitle{The thinnest cold HI clouds in the diffuse ISM?}

\shortauthors{S.\ Stanimirovi\'{c} \& C. Heiles}
\begin{document}

\title{The thinnest cold HI clouds in the diffuse interstellar medium?}

\author{Sne\v{z}ana Stanimirovi\'{c}, Carl Heiles}
\affil{Radio Astronomy Lab, UC Berkeley, 601 Campbell Hall,
Berkeley, CA 94720}
\email{sstanimi@astro.berkeley.edu, heiles@astro.berkeley.edu}

\begin{abstract}
We confirm and discuss recently discovered cold HI clouds with
column densities among the lowest ever detected.
The column densities of Cold Neutral Medium (CNM) towards 3C286 and
3C287 are $\sim10^{18}$ cm$^{-2}$, below an observational 
lower limit, and also below tiny-scale-atomic clouds 
detected by VLBI and time-variable profiles against pulsars. 
These column densities are close to the minimum
imposed by thermal evaporation. The fractions of the CNM
to total HI towards 3C286 and 3C287 are $\sim 4\%$ and $\lesssim2\%$, 
respectively. We discuss the CNM fraction and the CNM
clouds in relation to several theoretical models. \end{abstract}

\keywords{ISM: clouds --- ISM: structure --- radio lines: ISM}

\section{Introduction}

The Cold Neutral Medium (CNM) is easily studied with the 21-cm line
absorption because the HI opacity $\propto {N({\rm HI}) \over T}$, 
where $N({\rm HI})$
is the column density and $T$ the temperature. The CNM has been an
important subfield for the interstellar medium (ISM) in general and radio
astronomy in particular. Theoretically the CNM is understood as being one
of two thermal equilibrium states of the neutral medium 
\citep{Field69,McKee77a,Wolfire03}. 
\cite{McKee77a} summarized the expected properties for the 
CNM clouds: a constant density of 42 cm$^{-3}$, temperature
of 80 K, cloud size between 0.4 and 10 pc (with the mean value of 1.6
pc), the HI column density ranging between $6.5\times10^{19}$ and
$173\times10^{19}$ cm$^{-2}$, with the 
mean value of $27\times10^{19}$ cm$^{-2}$.

Although thermal properties of the CNM are very well understood
theoretically and observationally, its other aspects remain mysterious and
not well-studied, primarily because of a paucity of observational
knowledge. This observational paucity includes such very basics as cloud
shapes (morphology) and the column-density distribution; accompanying it
is the theoretical paucity involving production mechanism of the CNM,
the dynamic equilibrium between the CNM and its environment
(specifically, condensation, evaporation, and turbulent mixing), 
and the time scales for formation and destruction. 

Two important recent observational works are stimulating further work on
this topic. One is the Millennium Survey of 21-cm line absorption, of
which the latest paper (Heiles \& Troland 2005; HT05)  provides
statistical distributions of the fundamental parameters: column density,
temperature, turbulence, and magnetic field. The second is the detection
of reliable, very weak HI absorption lines towards three high-latitude
sources by \cite{Braun04} and Braun \& Kanekar (2005; BK05). 
These latest very sensitive HI
absorption observations were undertaken with the Westerbork radio
telescope, in the directions of four continuum sources, all located at
high Galactic latitude ($b\sim80$\degree) and relatively close to each
other.  The peak HI emission in these directions is very low, 2--5 K
only. For three out of four sources multiple absorption lines were
detected, with the peak optical depth of only 0.1 to 2\%. 

	BK05 discovered cold HI clouds with the lowest HI column
densities ever detected for cold interstellar clouds, more than  30 times
lower than what is expected for the smallest CNM clouds.  Fascinated by
the discoveries of BK05, and desiring to confirm their reality, we
repeated their observations for two sources, 3C286 and 3C287, with the
Arecibo telescope.  We easily confirmed findings by BK05 for 3C286.
Against 3C287 we see only a weak feature with too little signal/noise to
include in our discussion. 

	In this paper we report on the Arecibo observations of 3C286 and
3C287 and discuss implications of these findings. Our main aim is to
emphasize the existence of cold CNM clouds with the HI column densities
$\sim10^{18}$ cm$^{-2}$, to discuss these data in the context of
previous observational results, and to initiate discussion on the
possible origin of these clouds, as well as their importance in the ISM.
In Section 2 we briefly outline our observing and data processing
methods. Section 3 presents basic properties of cold HI clouds in the
directions of 3C286 and 3C287. 
A comparison of these clouds with CNM components
found in previous observational studies is given in Section 4.
In Section 5 we discuss several different
possibilities for the formation of these clouds.

\section{Observations and Data Processing}
\label{s:obs}

	The observations were conducted with  the Arecibo
telescope\footnote{The Arecibo Observatory is part of the National
Astronomy and Ionosphere Center, operated by Cornell University under a
cooperative agreement with the National Science Foundation.}. 
Several hours of observing time were granted at the discretion of the
Arecibo Observatory Director Sixto Gonzalez to evaluate the viability
of future very sensitive HI absorption measurements. 
The observing procedure was the same as in \cite{Heiles03a}. A specific
observing pattern, developed in \cite{Heiles03a}, was performed to
estimate the `expected' HI emission profile. This pattern generates one
on-source spectrum and 16 off-source spectra which are used to derive
first and second derivatives of the HI emission on the sky. 

The final velocity resolution is 0.16 \kms. The final 
rms noise level in the absorption spectra 
is $5\times10^{-4}$ over 0.5 \kms~channels (or $3\times10^{-4}$ over 
1 \kms~channels). 
After the data reduction, for both sources, 3C286 and 3C287,  we
have an HI absorption ($e^{-\tau(v)}$) and and an HI emission spectrum
($T_{\rm B}(v)$; note that this is the emission spectrum that would be
observed in the absence of the continuum source).

We next employed the technique by \cite{Heiles03a} to estimate the spin
temperature for the CNM clouds. This technique assumes that the CNM
contributes to  both HI absorption and emission spectra, while the
warm neutral medium (WNM) 
contributes only to the HI emission spectrum. The technique is based on
the Gaussian decomposition of both absorption and emission spectra, and
it  takes into account the fact that a certain fraction of the WNM gas
may be located in front of the CNM clouds resulting in only a portion of
the WNM being absorbed by the CNM.  
We note though that this technique may not be applicable for the case
when CNM occupies a solid angle significantly smaller than 
that of the Arecibo telescope beam (FWHM=3.5 arcmin), as is most 
likely the case with some CNM components we find in Section 4.3. 
In this case the observed absorption spectrum may not
correspond to the absorption that would be seen from all gas included
in the emission spectrum, and hence a direct comparison of the HI
absorption and emission spectra may not be a valid approach.
For those very small clouds that do not fill the beam our emission
intensity is underestimated, resulting in the derived spin temperature
being too low. Thus, we should regard our derived spin temperatures as
lower limits.
Another issue regarding this technique is the use of Gaussian
functions to represent the CNM absorption profiles.
\cite{Heiles03a} discussed pros and cons  of this approach in some detail.

\section{Results}
\label{s:results}

\begin{figure*}
\epsscale{1.7}
\plotone{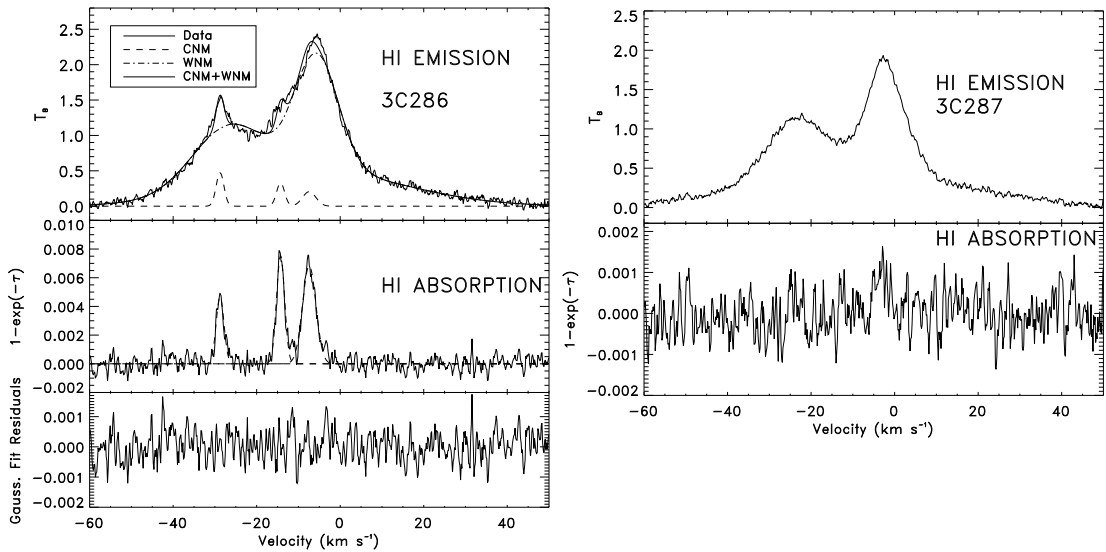}
\caption{\label{f:3c286} HI emission and absorption spectra for 3C286
(left) and 3C287 (right) obtained with the Arecibo telescope. 
For 3C286 three CNM components were detected.
For both sources the top panel shows the HI emission profile. In the
case of 3C286, separate contributions from the CNM and WNM to the HI
emission profile are shown in dashed and dot-dashed lines, respectively, while
the final, simultaneous fit is shown with the thick solid line. 
The middle panels shows the HI absorption spectra. 
For 3C286, three separate Gaussian functions were fitted (dashed
line). Note that 1-$\exp(-\tau) \approx \tau$ as the HI 
optical depths are very small. 
The bottom panel for 3C286 shows the residuals of the absorption
profile after the fitting procedure. Note that original spectra have
been box-car smoothed by three channels, the resulting rms noise in 
the resulting absorption spectra is $5\times10^{-4}$.}
\end{figure*}

In Fig.~\ref{f:3c286} we show HI emission and absorption spectra for
3C286 (on the left side) and 3C287 (on the right side). 
Three panels are shown for 3C286: ({\it Top:})
the HI emission spectrum (a main beam efficiency of 0.9 was used to
convert this spectrum from the antenna temperature units to the
brightness temperature scale) and separate contributions from the CNM
and WNM to the HI brightness temperature are shown with different lines;
the final (simultaneous) fit to the spectrum is also overlaid. ({\it
Middle:}) the HI absorption spectrum with fitted individual Gaussian
components. We detect three CNM components towards 3C286. 
({\it Bottom:}) the residuals of the absorption spectrum
for 3C286 after the fitting process. These random-looking residuals
demonstrate that the fit is perfect at the present level of signal/noise,
and that the Gaussian representation of CNM components is
valid in this case.
In the case of 3C287 we have only a marginal detection
at this stage.

\subsection{3C286}

	The HI absorption spectrum shows three distinct features, in
good agreement with BK05. We have fitted them with
three Gaussian functions centered at LSR velocities 
$-28.8$, $-14.3$, and $-7.4$ (Table 1).   The spin
temperature is well-constrained at the $10\%$ level for the first two
Gaussian components.
The contribution to the HI emission profile from the 
last CNM component appears to be very small, 
resulting in its spin temperature not being well constrained
($60\%$ level). The resultant HI column densities are directly proportional to
these temperatures and are given in Table 1 (column 7).  They are all
very low, $<1.5\times 10^{18}$ cm$^{-2}$. Table 1 also provides upper
limits to $N({\rm HI})$ assuming that the linewidths are thermal, in which
case the spin temperatures are equal to $T_{k,max}$. Even these are
small for the first two components, a few times $10^{18}$ cm$^{-1}$. 

We compare our results with BK05, who also found three distinct
absorption features at velocities and with peak optical depths very
similar to ours.  They fitted simultaneously HI emission and
absorption spectra as resulting from spherically symmetric, isobaric
clouds with an exponential HI volume density distribution. 
Assuming a representative spin temperature of 100 K, their estimated 
HI column densities for absorption features is
$0.4-8\times10^{18}$ cm$^{-2}$ \citep{Braun05}.

Using our derived temperatures from Table 1, the three
absorption components sum to the total estimated CNM column density
$3.9\times10^{18}$ cm$^{-2}$, while the total WNM column density is
$1.1\times10^{20}$ cm$^{-2}$. This means that the CNM comprises only 4\%
of the total HI column density in this direction. 

\subsection{3C287}

	Neither \cite{Dickey78} nor BK05 detected absorption features in
this direction, indicating that the peak optical depth must be $<0.01$.
We have only a marginal detection, at a 3-$\sigma$ level, of an
absorption feature at the LSR velocity of $-3.2$ \kms~with the peak
optical depth of only 10$^{-3}$. As the
signal/noise for this feature is too low we will place only an upper
limit for the CNM in this direction. Future observations will be able to
clarify this detection. 
Our upper limit on the peak optical depth (at a 5-$\sigma$ level) 
is $\sim 0.002$.
If we assume a hypothetical CNM component with
$\tau_{\rm max} < 0.002$,  $T_{\rm spin} = 100$ K, and FWHM$=3$ km s$^{-1}$,
then $N({\rm HI})_{\rm CNM} <1.2\times10^{18}$ cm$^{-2}$.
As the total WNM column density in this direction is $1.1\times10^{20}$
cm$^{-2}$, the CNM would comprise less than
$2\%$ of the total HI column density towards 3C287.


\begin{table*}
\footnotesize
\caption{Properties of CNM clouds in the directions of 3C286 and 3C287. }
\centering
\label{table1}
\begin{tabular}{lllllllll}
\noalign{\smallskip} \hline \hline \noalign{\smallskip}
Source   & l/b   & $\tau_{\rm max}$ & V$_{\rm LSR}^{a}$ & FWHM$^{a}$ & T$_{\rm
spin}$&T$_{\rm k,max}$ & $N({\rm HI})_{\rm CNM}$ & $N({\rm HI})_{\rm CNM}^{\rm upper}$ \\
         & (\degree/\degree)   &  & (\kms) & (\kms) & (K)& (K) &($10^{20}$ cm$^{-2}$) &($10^{20}$ cm$^{-2}$) \\
\hline
3C286 &          &       &       &  &   & &   &  \\
      &56.5/80.7 &$0.0047\pm0.0003$&$-28.8$&2.2  &$89\pm7$& 106 &$0.018\pm0.002$&$0.021$  \\
      &  & $0.0076\pm0.0003$&$-14.3$&2.3  &$37\pm4$& 115&    $0.013\pm0.002$   & $0.039$     \\
      &  & $0.0069\pm0.0002$&$-7.4$ &3.8  &$30\pm20$& 315 &  $0.015\pm0.008$   & $0.161$    \\
3C287 &  &       &       &  &       &                     &    \\
       &22.5/81.0 &$<0.002$&$\sim-3.2$ &3  &$100$& 200 &  $<0.012$ &$<0.023$ \\
\noalign{\smallskip} \hline \noalign{\smallskip}
\end{tabular}
\tablenotetext{a}{The uncertainy in the central velocity and FWHM is 
0.1 \kms\ for 3C286.}
\tablenotetext{}{The upper limit on HI column density, 
$N({\rm HI})_{\rm CNM}^{\rm upper}$, was derived using $T_{\rm k,max}$
instead of $T_{\rm spin}$.}
\end{table*}


\section{Discussion: observational issues}
\label{s:discussion}

\subsection{A new population of interstellar clouds ?}

	The detected peak optical depths here and in BK05
are among the lowest ever detected  for Galactic CNM clouds. 
They appear to represent a new population of CNM clouds that is 
not represented by the existing observational data and summaries.

For 3C286 and 3C287 the estimated HI column densities are all
$<1.6\times 10^{18}$ cm$^{-2}$, more than 30 times lower than 
what is traditionally expected for the smallest CNM clouds 
\citep{McKee77a}. Clearly, the CNM clouds in
the directions of 3C286 and 3C287 have very low HI column density, and
as shown in the previous section, these CNM clouds comprise only
$<2$--4\% of the total measured HI column density.

	But how unusual are these low CNM columns? In the most recent HI
absorption survey with the Arecibo telescope,  \cite{Heiles03b} found
that the median HI column density for CNM clouds is $5\times10^{19}$
cm$^{-2}$ (note that the sensitivity of this survey is $\Delta N({\rm HI})\sim
10^{18}$ cm$^{-2}$). HT05 provided statistical distributions of the
fundamental parameters: HI column density, temperature, turbulence, and
magnetic field. They found that the observed probability density
function (PDF) of $N({\rm HI})$, for sources primarely at high Galactic latitudes, 
is well-approximated by $N^{-1}$ between the limits given by: 

\begin{mathletters} \label{Nobsobsnumbers} 
\begin{equation}
{\Nperp}_{20,min} = 0.026 _{- 0.010}^{+0.019} \ {\rm cm}^{-2}
\end{equation} 

\begin{equation} {\Nperp}_{20,max} = 2.6  _{- 1.2}^{+1.9}
\ {\rm cm}^{-2}, 
\end{equation} 
\end{mathletters}
where the subscript 20 means units of $10^{20}$ cm$^{-2}$. 

	HT05 regarded the lower limit (${\Nperp}_{20,min}$) to be real,
not an observational result imposed by the lack of sensitivity. 
However, the presence of three weak absorption components towards
3C286---one of HT05's sightlines---shows that their sensitivity estimate was
optimistic.  HT05's PDF of equation \ref{Nobsobsnumbers} certainly
applies to their sources with strong absorption lines which have
received a large amount of  integration time for measuring Zeeman
splitting. But it appears that there is a new population of CNM
components that is unrepresented in their survey.

\subsection{The CNM fraction}

McKee and Ostriker (1977) predicted the CNM to form $\sim95\%$
of the total HI. Observers have traditionally found smaller
column-density fractions, for example \cite{Heiles03b} found that
the majority of sightlines have the CNM column density fraction $<30\%$,
while the WNM fraction is $\sim61\%$. This WNM column density fraction
translates to a volume fraction of $\sim0.5$.
\cite{Wolfire03} have updated the theory of ISM neutral phases and slightly 
refined values for the WNM pressure and density; 
these refinements bring the observed WNM volume fraction to 0.8.
However, sightlines towards 3C286 and 3C287 have much smaller 
fractions of the CNM column density, $\sim 4\%$ and $\lesssim 2\%$, 
respectively. 
These echo the small fractions found by HT05, who found 19 (out of 79) 
lines of sight with no detectable CNM; 3C286 was one of these, 
so perhaps we should take their non-detections to indicate a 
possible representative upper limit of a few percent for 
this CNM/total HI ratio. 

These small CNM fractions might be a problem for the theory.
Alternatively, the particular sightlines having small CNM fractions
might lie in special regions affected by energetic processes that have
temporarily destroyed the CNM. 
In fact, sightlines towards 3C286 and 3C287 pass through
the North Galactic Pole region where \cite{Kulkarni85a} already noticed 
that about 50\% of HI is infalling towards the plane,
while warm HI occupies about 40\%.
This question needs further investigation.

\subsection{Are low column density clouds related to the tiny-scale
atomic structure?}

	A second population of low-column CNM clouds are those often
referred to as the tiny-scale atomic structure (TSAS). The TSAS size
scale is inferred from time variability of HI  absorption profiles
against pulsars or VLBI imaging of  extra-galactic continuum sources and
ranges from a few AU to a few hundreds of AU.  
TSAS typically has $N({\rm HI})$ somewhat
higher than our Table 1 components, namely from $\sim3\times10^{18}$ to
$\sim2\times10^{19}$ cm$^{-2}$ \citep{Heiles97}. Recently,
\cite{Stanimirovic05} found persistent variations in the HI optical
depth profiles of PSR B1929+10 which indicate structure in the cold HI
with $N({\rm HI})\sim2\times10^{18}$ cm$^{-2}$.  With these column
densities and small sizes, the TSAS must be overpessured with respect to
most of the ISM. Thus TSAS features are thought to be very dense and
over-pressured, with $n({\rm HI})$ and $P$ reaching values up to
$\sim10^{4}$ cm$^{-3}$  and $\sim10^{6}$ K cm$^{-3}$, respectively.

	If we assume that our low column density CNM clouds towards
3C286 and 3C287 are similarly overpressured, then the implied size scale
is even smaller than the typical TSAS scale because our column
densities are somewhat smaller. Alternatively, if our CNM components are
 at the standard ISM pressure, with $nT \sim 3000$ cm$^{-3}$ K
\citep{Jenkins01}, then their HI volume densities are $n \sim 20-100$
cm$^{-3}$ and the implied size scales  
are ${N({\rm HI}) \over n({\rm HI})} \sim
800-4000$ AU.  Even these larger sizes are not too much higher than the
inferred sizes for much of the TSAS. 

	Another indication that the low column density clouds could be
related to TSAS comes from direct interferometric imaging. Over a range
of velocities BK05 found compact emission clumps with a FWHM of 1--2
\kms\ and an intrinsic size of 30 arcsec; at an assumed distance of 100
pc this corresponds to  a plane-on-the-sky size of about 3000 AU. 
This morphology is distinctly different from the clumpy sheet-like model
of the CNM suggested by \cite{Heiles03b}, and may point again to a
distinctly different origin of these clouds. The compact clumps at
apparently random locations may be suggestive  of fluctuations in the
distribution of HI optical depth over a wide range of spatial scales,
rather than the presence of distinct physical entities. If the low
column density HI clouds and TSAS features have a common origin, then
TSAS is significantly more abundant in the ISM, and with  significantly
lower optical depths, than what is expected theoretically
if the TSAS is simply the low-size-scale extension of the interstellar
turbulence spectrum as discussed by \cite{Deshpande00a}. 

	Perhaps the TSAS is characterized by a range of column densities
and pressures, with the classical absorption observations presented here
sampling  the low-density end and the pulsar/VLBI observations, which
are less sensitive, the (possibly much rarer)  high-density end.

\section{Discussion: theoretical considerations}

We discuss some current theoretical approaches which can lead to the
production of cold clouds in the ISM with column densities similar to
what is found observationally. These comprise two general classes:
clouds that are long-lived and can be characterized as semi-permanent
structures, and transient clouds.

\subsection{Semi-permanent clouds}

\subsubsection{Evaporation versus condensation} 

In the well-accepted theory of \cite{McKee77a} the CNM resides within
warmer phases.
However, it is expected that different phases are
separated by an interface region through which heat flows between the
two phases. The interface is a surface of transition where, for small
CNM clouds, the CNM evaporates into the warmer medium and, for large CNM
clouds, the warmer medium condenses onto the CNM.  
The classical evaporation theory
\citep{McKee77b}  predicts a critical radius for CNM clouds at which
radiative losses balance the conductive heat input and at which the
cloud neither evaporates nor condenses. Clouds smaller than the critical
radius evaporate, while clouds larger than the critical radius accrete
by condensation of the surroundings. 
\cite{McKee77b} estimated the 
characteristic scale length for different
temperature layers in the interface region.
From there, in the case of the least intrusive
interface, CNM to WNM with the intercloud temperature of $10^{2}-10^{4}$
K,  the expected lower limit on the column density of the interface region  is
about $N({\rm HI})\sim3\times10^{17}$ cm$^{-2}$. 
Higher intercloud temperatures are
more intrusive, for example for $T\sim10^{6}$ K the interface
region would have several temperature layers with different column
densities, adding up to the total column density $>2\times10^{18}$
cm$^{-2}$, mainly ionized though.

In the case of the least intrusive interface with the WNM, the cloud
critical radius ($R_{\rm rad}$) times the density of the intercloud
region (we assume $n_{\rm WNM}=0.5$ cm$^{-3}$) is about $5\times10^{16}$
cm$^{-2}$ (from Fig. 2 in McKee \& Cowie 1977). The expected critical
radius is then $R_{\rm rad}\sim6\times10^{3}$ AU, meaning that clouds
smaller  than this size evaporate, while CNM clouds larger than $R_{\rm
rad}$ accrete the WNM by condensation. 
These small, evaporating clouds are surprisingly long-lived, 
typical evaporation time scale is
of order of $10^{6}$ yr (we have assumed the mass-loss
rate of $1.2\times10^{23}$ gr yr$^{-1}$ and $n_{\rm CNM}=30$ cm$^{-3}$,
and used equation 47 in  McKee \& Cowie 1977). 

The estimated sizes for the low-$N$(HI) clouds in Section 4.3
are all below the critical cloud radius even in the case of the
least-intrusive WNM. These CNM clouds, protected from ionization by
large WNM envelopes, should be evaporating but 
over long time scales, and hence could be common in the ISM.

\subsubsection{Condensation of WNM into CNM induced by turbulence ?}

Recent numerical simulations by \cite{Audit04} show 
that a collision of incoming turbulent flows can initiate
condensation of WNM into cold neutral clouds.
This model hence naturally explains a formation of cold clouds
at high Galactic latitudes.
In the simulations, a collision of incoming WNM streams creates
a thermally unstable region of higher density 
and pressure but lower temperature, which further fragments into cold
structures. The thermally unstable gas has a filamentary morphology and 
its fragmentation into cold clouds is promoted and controlled by turbulence.
Typical properties of cold clouds formed in the simulations are:
$n\sim50$ cm$^{-3}$, $T\sim80$ K, $R\sim0.1$ pc. Also, 
a lot of small clouds, with sizes reaching the numerical 
resolution of 0.02 pc, were found.

These cold clouds are thermally stable and long-lived, 
in the case of stronger turbulence they are more roundish, 
while a weaker turbulence 
is responsible for more elongated morphology of cold clouds.
The simulations show that the fraction of cold gas ranges from 10\% in a
strong turbulent case, to about 30\% in a weak turbulent case. These
fractions are larger than our observed fractions; perhaps further
refinement of the models can produce better agreement. 
In fact, it is known that the CNM fraction and the CNM column density
depend also on the input mean thermal pressure and the input velocity of
colliding flows which are free parameters in the simulations and
could be better constrained from HI observations
(P. Hennebelle, private communication).

These simulations show that a collision of turbulent WNM  streams is
capable of producing a large number of  small CNM clouds with low column
densities,  and not confined to the Galactic plane. The CNM clouds are
thermally stable and embedded in large, unstable WNM filaments.  The
abundance of cold clouds, as well as their properties,  depend heavily
on the properties of the underlaying turbulent flows. In a similar
approach, \cite{Koyama02} showed that a thermally unstable 
shock-compressed layer can also fragment into small, cold, turbulent
condensations.

\subsection{Transient clouds}

	A CNM cloud is normally pictured as an entity in 
quasi-equilibrium with its surroundings.  A contrasting viewpoint,
stimulated by the recent flurry of numerical simulations of interstellar
turbulence, envisions the clouds as dynamic entities that are constantly
changing in response to the turbulent ``weather'' (for a review, see
\cite{MacLow04}). For example, Vazquez-Semadeni, Ballesteros-Paredes, \&
Rodriguez (1997) performed simulations that produced a lot of clouds
with very small sizes and low column densities ($<10^{19}$ cm$^{-2}$).
These clouds are out of equilibrium and probably very transient. 

In MHD simulations by \cite{Semadeni97} cold clouds form  at the
interfaces of expanding shells.  Their typical size is $\sim1$--200 pc
(see Fig. 5 in Vazquez-Semadeni, Ballesteros-Paredes, \& Rodriguez
1997), and for any given  size the mean volume density of clouds ranges
from $\sim2$ to 60 cm$^{-3}$.   Clouds are found to  have irregular and
filamentary shapes. The cloud column density ranges from $10^{19}$ to
$10^{20}$ cm$^{-2}$, while a typical FWHM is from 2 to 15
\kms. 

Qualitatively different is the interaction of shocks with
pre-existing clouds, as opposed to clouds forming within shocks.
Nakamura, Klein, McKee \& Fisher (private communication) find 
that when a supernova shock
hits a pre-existing cloud, the cloud is torn up into small pieces (`shreds')
with a velocity dispersion of several \kms. Some of the predicted
properties of these `shreds' appear close to what we observe. Further
advances in this area are eagerly awaited for more detailed comparison
with observations. 

In summary, several properties predicted by 
numerical simulations do not match
our observations very well, others do. These theoretical efforts are in
their early stages and  are  developing rapidly. We can expect rapid
evolution in the future.

\section{Summary and Conclusions}

Recently, Braun \& Kanekar (2005) discovered several cold HI clouds at
high Galactic latitudes with small HI optical depths. We have
confirmed their detections in the direction of 3C286
and added an additional weak upper limit to
this set. We have found the absorption components to be well-represented by
Gaussian functions and have derived their spin temperatures 
and HI column densities. These
clouds have HI column densities among the lowest ever detected for the CNM
features and could represent a new population of interstellar clouds. 
We have compared them with the TSAS detected by time-variable pulsar
absorption lines and VLBI and concluded that they might be of the same
general class. These observational techniques used for detecting TSAS
might highlight the high-volume density and pressure 
members of this class, while the more traditional 
emission/absorption observations highlight the lower-volume 
density/pressure members.

The sightlines towards 3C286 and 3C287 have very low CNM fractions.
Similarly low fractions have been found for 24\% of continuum sources
in the large HI absorption line survey by \cite{Heiles03a}. 
Such low CNM fractions are not theoretically predicted. 
This might be either a problem with the
theories or might indicate that some sightlines are special, traversing
regions from which CNM has been removed perhaps by energetic processes.

	We have investigated briefly three different theoretical
approaches for the production of cold, low column density clouds in the
ISM. Two of these envision the CNM clouds as semi-permanent entities. 
The third envisions the clouds as
transitory structures. CNM clouds may be formed by cooling and
condensation in colliding turbulent streams of WNM, whereby cold clouds
are stable discrete entities surrounded by thermally unstable WNM
filaments. Alternatively, many numerical simulations envision cold
clouds being transient and constantly changing phenomena in the
turbulent ISM. Theory and observations are not in full accord regarding
the fraction of CNM gas and its column densities. 
Further work in both observations and theory is called for. 

	Observationally, it is essential to quantify how common these
clouds are in the ISM and whether they are related to some local events,
such as stellar winds or large scale atomic flows. We need to increase
the statistical sample to establish the probability density function of
the column density in the vicinity of its current lower cutoff where
most of the low-$N$(HI) clouds are found. In addition we need to more firmly
determine the relationship between the CNM components found here and
TSAS.

\begin{acknowledgements}
We are grateful to telescope operators at the Arecibo Observatory for
their help in conducting these observations, 
particularly to William Torres and Norberto Despiau. We
would also like to thank Hector Hernandez and Sixto Gonzalez for prompt 
scheduling of this project. It is a pleasure to acknowledge
stimulating discussions with Patrick Hennebelle, Miller Goss, 
Joel Weisberg, Nissim Kanekar, Robert Braun and Chris McKee. 
SS is indebted to Chris McKee for pointing out an error in
an earlier version of this manuscript.
We also thank an anonymous referee for valuable suggestions.
Support by NSF grants  AST-0097417 and AST-9981308 is gratefully acknowledged.
\end{acknowledgements}

\label{lastpage}
\end{document}